\documentclass{article}
\usepackage{spconf,amsmath,graphicx}
\usepackage{amsfonts}
\usepackage{subcaption}
\usepackage{tikz}
\usetikzlibrary{calc,chains,shapes,positioning}
\usepackage{pgfplots}
\usepackage{phaistos}
\usepackage{float}
\usepackage{phaistos}
\usepackage{pgfplots}
\usepackage{cite}
\usepackage{mathdots}
\usepackage{scalerel}
\usepackage{enumitem}
\usepackage[subtle,title=normal,sections=normal,margins=normal,bibliography=normal,mathdisplays=normal,mathspacing=normal]{savetrees}
\usepackage{balance}
\usepackage{epstopdf}

\title{Dereverberation in Acoustic Sensor Networks using weighted\\ prediction error with Microphone-dependent Prediction Delays}
\name{Anselm Lohmann$^{1}$, Toon van Waterschoot$^{2}$, Joerg Bitzer$^{3}$, Simon Doclo$^{1}$ \thanks{This work has received funding from the European Union’s Horizon 2020 research and innovation programme under the Marie Skłodowska-Curie grant
agreement No. 956369 and the ERC Consolidator Grant: SONORA (no. 773268), and from the Deutsche Forschungsgemeinschaft (DFG, German Research Foundation) under Germany's Excellence Strategy - EXC 2177/1 - Project ID 390895286.}}
\address{$^{1}$University of Oldenburg, Dept. of Medical Physics and Acoustics, Oldenburg, Germany \\
$^{2}$KU Leuven, Department of Electrical Engineering (ESAT-STADIUS), Leuven, Belgium\\
$^{3}$Fraunhofer IDMT, Project Group Hearing, Speech and Audio Technology, Oldenburg, Germany \\
{\tt anselm.lohmann@uni-oldenburg.de}
}

\begin{document}
\ninept
\maketitle
\begin{abstract}
In the last decades several multi-microphone speech dereverberation algorithms have been proposed, among which the weighted prediction error (WPE) algorithm. In the WPE algorithm, a prediction delay is required to reduce the correlation between the prediction signals and the direct component in the reference microphone signal. In compact arrays with closely-spaced microphones, the prediction delay is often chosen microphone-independent. In acoustic sensor networks with spatially distributed microphones, large time-differences-of-arrival (TDOAs) of the speech source between the reference microphone and other microphones typically occur. Hence, when using a microphone-independent prediction delay the reference and prediction signals may still be significantly correlated, leading to distortion in the dereverberated output signal. In order to decorrelate the signals, in this paper we propose to apply TDOA compensation with respect to the reference microphone, resulting in microphone-dependent prediction delays for the WPE algorithm. We consider both optimal TDOA compensation using crossband filtering in the short-time Fourier transform domain as well as band-to-band and integer delay approximations. Simulation results for different reverberation times using oracle as well as estimated TDOAs clearly show the benefit of using microphone-dependent prediction delays.
\end{abstract}
\section{Introduction}
\label{sec:intro}
When recording a speech source using microphones inside a room, reverberation due to acoustic reflections may degrade the quality and intelligibility of the recorded speech. While early reflections may be beneficial, late reverberation typically reduces both speech intelligibility as well as automatic speech recognition performance \cite{Beutelmann2006,YoshiokaASR2012}. Hence, effective dereverberation is required for many speech communication applications, such as voice-controlled systems, hearing aids and hands-free telephony \cite{HabetsNaylor2018,Cauchi2015,Braun2018,Dietzen2019,Li2019,Williamson2017,Lemercier2022,Nakatani2010,Jukic2015,Wung2020,Witkowski2021,Huang2022}. A popular blind dereverberation algorithm is the weighted prediction error (WPE) algorithm \cite{Nakatani2010,Jukic2015,Wung2020,Witkowski2021,Huang2022}, which is based on multi-channel linear prediction (MCLP). WPE performs dereverberation by first estimating the late reverberant component in a reference microphone from the delayed reverberant microphone signals and then subtracting the estimate from the reference microphone signal. Several variants of the WPE algorithm have been proposed, e.g., aiming at controlling sparsity of the dereverberated output signal in the time-frequency domain \cite{Jukic2015,Witkowski2021}.  
\begin{figure}[t!]
\hspace{1.28cm}\includegraphics[width=0.7\columnwidth]{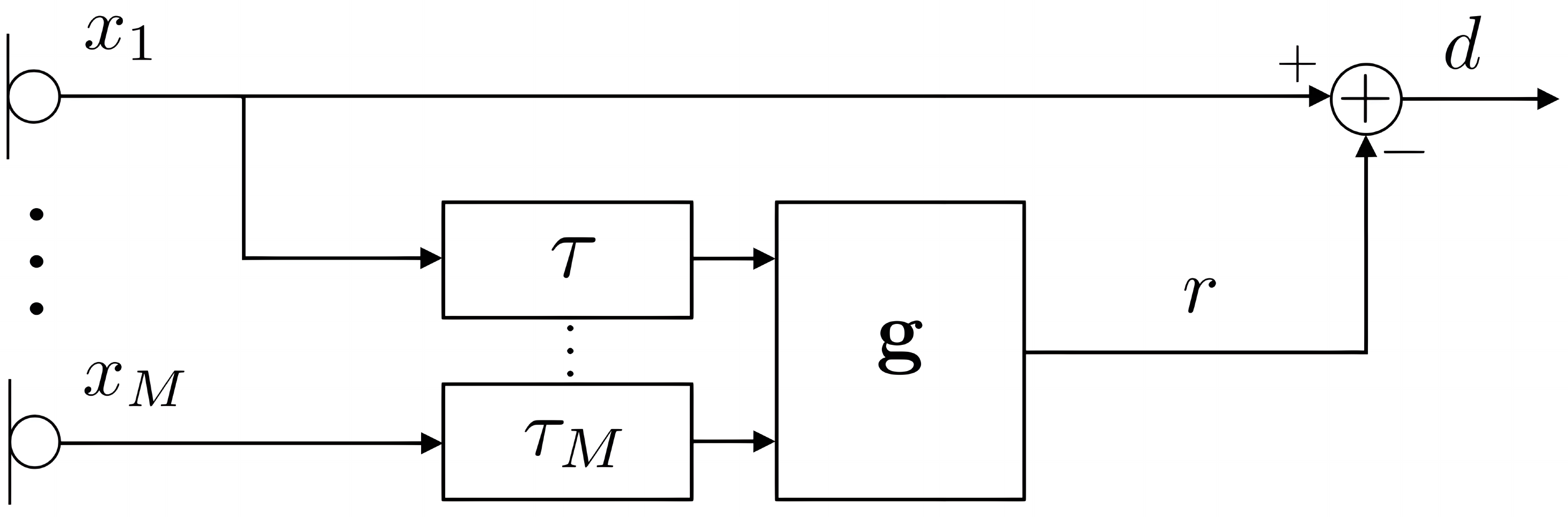}
\caption{WPE with microphone-dependent prediction delays}
\label{fig:block_diag}
\vspace{-0.6cm}
\end{figure}%
\par The delay, called prediction delay, plays an important role in WPE and is introduced to reduce the correlation between the prediction signals and the direct component in the reference microphone signal, hence aiming to preserve the direct component and early reflections \cite{Nakatani2010,Jukic2015}. The prediction delay determines the trade-off between the amount of residual reverberation and the distortion in the desired component. It is typically chosen according to the autocorrelation of speech and is therefore often chosen microphone-independent. Whereas this may be a good choice for compact arrays with closely-spaced microphones, in acoustic sensor networks with spatially distributed microphones large inter-microphone distances may exist, leading to large and diverse time-differences-of-arrival (TDOAs) of the speech source between microphones. If uncompensated for, the effective prediction delay in each microphone may be sub-optimal, leading to distortion or excess reverberation in the desired component. \par Aiming at performing TDOA compensation with respect to the reference microphone in the short-time Fourier transform (STFT)-domain, in this paper we propose to use microphone-dependent prediction delays in the WPE algorithm. We perform either optimal TDOA compensation with non-integer delays or coarse TDOA compensation with integer delays. In the STFT-domain, optimal TDOA compensation needs to be implemented using crossband filtering. However, since crossband filters introduce additional computational complexity and using more crossband filters does not necessarily imply lower reconstruction error in subbands \cite{Avargel2007}, we also propose using a band-to-band approximation. Simulation results for an acoustic sensor network with spatially distributed microphones using different reverberation times as well as estimated and oracle TDOAs show the proposed microphone-dependent prediction delays outperform the microphone-independent prediction delay in the WPE algorithm, where the best dereverberation performance is achieved using non-integer delays with crossband filtering closely followed by the computationally less complex band-to-band approximation.

\section{Signal Model}
\label{sec:format}
We consider a single speech source captured by a set of $M$ microphones in a reverberant room. Similarly as in \cite{Nakatani2010,Jukic2015,Witkowski2021}, we don't consider additive noise in this paper. In the STFT-domain, let $s(k,n)$ denote the clean speech signal with $k \in \{1,...,K\}$ the subband index and $n \in \{1,...,N\}$ the time frame index. The reverberant signal at the $m$-th microphone $x_m(k,n)$ can be written as
\begin{equation} 
    x_m(k,n) = \sum_{l = 0}^{L_h - 1}h_m(k,l)s(k,n-l) + e_m(k,n),
    \label{eq:sigmodel_stft}
\end{equation}
where $h_m(k,n)$ denotes the subband convolutive transfer function with length $L_h$ between the speech source and the $m$-th microphone, and $e_m(k,n)$ denotes the subband modelling error. Without loss of generality, we define the first microphone as the reference microphone. Assuming the error term $e_m(k,n)$ in \eqref{eq:sigmodel_stft} can be disregarded, the dereverberation problem as is depicted in Fig. \ref{fig:block_diag} can be formulated as
\begin{equation}
    d(k,n) = x_1(k,n) - r(k,n).
    \label{eq:sig_model}
\end{equation}
The desired component $d(k,n) = \sum_{l = 0}^{L_d - 1}h_1(k,l)s(k,n-l)$ consists of the direct path and early reflections in the reference microphone signal $x_1(k,n)$, where $L_d$ denotes the temporal cut-off between early and late reflections. The undesired component $r(k,n) = \sum_{l = L_d}^{L_h - 1}h_1(k,l)s(k,n-l)$, which we aim to estimate, is the late reverberant component in the reference microphone signal $x_1(k,n)$. Using the MCLP model \cite{Nakatani2010}, the undesired late reverberant component $r(k,n)$ can be written as the sum of delayed filtered versions of all reverberant microphone signals, i.e.
\begin{equation}
r(k,n) = \sum_{m = 1}^{M}\sum_{l=0}^{\tilde{L}_g - 1}\tilde{g}_m(k,l)x_m(k,n - \tau - l),
\label{eq:conv_MCLP}
\end{equation}
where $\tilde{g}_m(k,n)$ denotes the $m$-th prediction filter of length $\tilde{L}_g$ and $\tau = L_d$ denotes the (microphone-independent) prediction delay. \\
The prediction delay aims at reducing the correlation between the prediction signal $x_m(k,n-\tau)$ and the desired component $d(k,n)$. It is typically chosen according to the autocorrelation of speech and therefore often microphone-independent.
In this paper we propose to generalise the MCLP model in \eqref{eq:conv_MCLP} to allow for a microphone-dependent prediction delay $\tau_m$ (see Fig. \ref{fig:block_diag}), i.e.
\begin{equation}
\boxed{r(k,n) = \sum_{m = 1}^{M}\sum_{l=0}^{L_g - 1}g_m(k,l)x_m(k,n - \tau_m - l),}
\label{eq:new_MCLP}
\end{equation}
where $g_m(k,l)$ denotes the $m$-th prediction filter of length $L_g$. Using \eqref{eq:new_MCLP}, the signal model in \eqref{eq:sig_model} can be rewritten in vector notation  as
\begin{equation}
    \mathbf{d}(k) = \mathbf{x}_1(k) - \mathbf{X}_{\boldsymbol{\tau}}(k)\mathbf{g}(k),
    \label{eq:new_vec_MCLP}
\end{equation}
with
\begin{equation}
\mathbf{d}(k) = \begin{bmatrix} 
d(k,1)&\cdots & d(k,N)
\end{bmatrix}^{T} \in\mathbb{C}^{N},
\end{equation}
\begin{equation}
\mathbf{x}_1(k) = \begin{bmatrix} 
x_1(k,1)&\cdots & x_1(k,N) 
\end{bmatrix}^{T} \in\mathbb{C}^{N},
\end{equation}
where $N$ denotes the number of time frames.
The multi-channel delayed convolution matrix $\mathbf{X}_{\boldsymbol{\tau}}(k)$ in \eqref{eq:new_vec_MCLP} is defined as
\begin{equation}
\mathbf{X}_{\boldsymbol{\tau}}(k) = \begin{bmatrix} 
\mathbf{X}_{\tau_1}(k)&\cdots & \mathbf{X}_{\tau_{M}}(k)
\end{bmatrix}\in\mathbb{C}^{N\times ML_g},
\end{equation} 
where $\mathbf{X}_{\tau_m}(k) \in\mathbb{C}^{N\times L_g}$ is the convolution matrix of $\mathbf{x}_m(k)$ delayed by $\tau_m$ frames with $\tau_1 = \tau$ denoting the prediction delay in the reference microphone and $\mathbf{g}(k) \in \mathbb{C}^{ML_g}$ is the stacked vector of all prediction filter coefficients $g_m(k,n)$. The problem of speech dereverberation, i.e. estimation of the desired component $\mathbf{d}(k)$, is now reduced to estimating the filter $\mathbf{g}(k)$ predicting the undesired late reverberation.
\section{WPE Algorithm}
\label{sec:pagestyle}
To estimate the prediction filter $\mathbf{g}(k)$, it has been proposed in \cite{Nakatani2010} to model the desired component $d(k,n)$ using a time-varying Gaussian (TVG) model. This is equivalent to modelling the desired component in each time-frequency bin by means of a zero-mean circular Gaussian distribution with time-varying variance $\lambda(k,n)$, i.e.
\begin{equation}
    \mathcal{N}_{\mathbb{C}}(d(k,n);0,\lambda(k,n))=\frac{1}{\pi\lambda(k,n)}e^{-\frac{\lvert d(k,n) \rvert^2}{\lambda(k,n)}},
    \label{eq:TVG}
\end{equation}
where the variance $\lambda(k,n)$ is an unknown parameter which needs to be estimated. Since the TVG model does not assume any dependency across frequencies or time frames, the likelihood function is given by
\begin{equation}
    \mathcal{L}(\mathbf{g}(k),\boldsymbol{\lambda}(k)) = \prod_{n=1}^{N}\mathcal{N}_{\mathbb{C}}(d(k,n);0,\lambda(k,n)),
\end{equation}
with
\begin{equation}
\boldsymbol{\lambda}(k) = \begin{bmatrix} 
\lambda(k,1)&\cdots & \lambda(k,N)
\end{bmatrix}^{\mathrm{T}} \in\mathbb{R}^{N}.
\end{equation}
The prediction filter $\mathbf{g}(k)$ and the variances $\boldsymbol{\lambda}(k)$ can then be estimated by maximising the log-likelihood function, which is equivalent to solving the following optimisation problem \cite{Nakatani2010, Jukic2015}
\begin{equation}
    \min_{\boldsymbol{\lambda}(k),\mathbf{g}(k)} \sum_{n=1}^{N}\left(\frac{\lvert d(k,n)\rvert^2}{\lambda(k,n)} + \text{log}\pi\lambda(k,n)\right).
    \label{eq:WPE_optim}
\end{equation}
Since no closed-form solution exists for the joint optimisation problem in \eqref{eq:WPE_optim}, it has been proposed in \cite{Nakatani2010} to use an alternating optimisation procedure,  where two simpler sub-problems are solved in an iterative manner. More in particular, in each iteration the cost function in \eqref{eq:WPE_optim} is first minimized with respect to the prediction vector $\mathbf{g}(k)$, assuming that the variances $\boldsymbol{\lambda}(k)$ are fixed to the values from the previous iteration. Using these values for the prediction vector $\mathbf{g}(k)$, the cost function in \eqref{eq:WPE_optim} is then minimized with respect to the variances $\boldsymbol{\lambda}(k)$. At a given iteration $i$, with index $k$ omitted, the estimated prediction filter and the estimated variances are given by
\begin{equation}
    \hat{\mathbf{g}}^{(i+1)} = \left(\mathbf{X}_{\boldsymbol{\tau}}^{H}\mathcal{D}^{-1}_{\hat{\boldsymbol{\lambda}}^{(i)}}\mathbf{X}_{\boldsymbol{\tau}}\right)^{-1}\mathbf{X}_{\boldsymbol{\tau}}^{H}\mathcal{D}^{-1}_{\hat{\boldsymbol{\lambda}}^{(i)}}\mathbf{x}_1,
\end{equation}
\begin{equation}
    \hat{\boldsymbol{\lambda}}^{(i+1)} = \lvert\hat{\mathbf{d}}^{(i+1)}\rvert^2 = \lvert \mathbf{x}_1 - \mathbf{X}_{\boldsymbol{\tau}}\mathbf{g}^{(i+1)}\rvert^2,
    \label{eq:variance_update}
\end{equation}
where $\mathcal{D}_{\hat{\boldsymbol{\lambda}}^{(i)}} = \text{diag}(\boldsymbol{\hat{\lambda}}^{(i)})$ and $\hat{\mathbf{d}}^{(i+1)}$ is the estimated desired component calculated using \eqref{eq:new_vec_MCLP} with the $\lvert.\rvert$ and $(.)^2$ operators applied element-wise.\par
In \cite{Jukic2015}, an extension to the TVG model was proposed by including a hyperprior with a sparsity-promoting parameter $p$, aimed at better describing the sparse nature of speech in the STFT-domain. It was shown in \cite{Jukic2015} that the solution for the variances in \eqref{eq:variance_update} simply changed to
\begin{equation}
    \hat{\boldsymbol{\lambda}}^{(i+1)} = \lvert\hat{\mathbf{d}}^{(i+1)}\rvert^{2-p}.
    \label{eq:sparsity_p}
\end{equation}
\section{Microphone-dependent Prediction Delays}
In acoustic sensor networks with spatially distributed microphones, large and diverse TDOAs of the speech source between the microphones may exist. If uncompensated for, the effective prediction delays in the microphones may be suboptimal, leading to distortion or excess reverberation in the WPE output signal (see evaluation in Section 5). In this section we propose to perform TDOA compensation using the microphone-dependent prediction delays in \eqref{eq:new_MCLP}. We perform either optimal TDOA compensation with non-integer microphone-dependent prediction delays or coarse TDOA compensation with integer microphone-dependent prediction delays. In Section 4.1 we first define the problem in the time-domain. It can be shown that the optimal TDOA compensation filter in the time-domain can be written as a combination of integer frame-delays and crossband filters in the STFT-domain (see Section 4.2). To reduce computational complexity, we also propose using a band-to-band approximation (Section 4.3) or using integer frame-delays (Section 4.4).

\subsection{TDOA compensation in time-domain}
Let $\text{TDOA}_{m}$ be the TDOA between microphone $m$ and the reference microphone. This TDOA can be decomposed as
\begin{equation}
    \text{TDOA}_{m} = \delta_m^{\text{frame}}\times L_\text{shift} + \delta_m^{\text{samp}} + \delta_m^{\text{frac}},
\end{equation}
with the integer frame-delay $\delta_m^{\text{frame}}$, the integer delay $\delta_m^{\text{samp}}$ and the fractional delay $\delta_m^{\text{frac}}$ defined as
\begin{equation}
\delta_m^{\text{frame}} = \lfloor\frac{\text{TDOA}_{m}}{L_\text{shift}}\rceil,
\end{equation}
\begin{equation}
\delta_m^{\text{samp}} = \lfloor\text{TDOA}_{m} - \delta_m^{\text{frame}}\times L_\text{shift}\rfloor,
\end{equation}
\begin{equation}
\delta_m^{\text{frac}} = \text{TDOA}_{m} - \delta_m^{\text{frame}}\times L_\text{shift} - \delta_m^{\text{samp}},
\end{equation}
where $L_\text{shift}$ denotes the STFT frame shift, $\lfloor.\rceil$ denotes the operator for nearest integer rounding and $\lfloor.\rfloor$ denotes the operator for lower integer rounding. The optimal time-domain TDOA compensation filter $\underline{u}'_{m}[t]$ can be written as
\begin{equation}
    \underline{u}'_{m}[t] = \delta[t - \delta_m^{\text{frame}}\times L_{\text{shift}}]*\delta[t - \delta_m^{\text{samp}}]*\text{sinc}[t - \delta_m^{\text{frac}}],
    \label{eq:time_domain}
\end{equation}
where $\delta[t]$ denotes a unit impulse with discrete-time index $t$ and $*$ denotes convolution. It should be noted that $\delta[t - \delta_m^{\text{frame}}\times L_{\text{shift}}]$ can be implemented using integer frame-delays in the STFT-domain, whereas $\underline{u}_{m}[t] = \delta[t - \delta_m^{\text{samp}}]*\text{sinc}[t - \delta_m^{\text{frac}}]$ needs to be implemented using crossband filters in the STFT-domain.
\subsection{Non-integer prediction delays using crossband filtering}
It was shown in \cite{Avargel2007} that any time-domain filter, e.g. $\underline{u}_m[t]$, can be implemented in the STFT-domain using crossband filters
\begin{equation}
    u_m(k,k',n) = \{\underline{u}_m[t] * \phi_{k,k'}[t]\}\lvert_{t = nL_{\text{{shift}}}},
    \label{eq:opt1}
\end{equation}
where $u_m(k,k',n)$ denotes the crossband filter with crossband index $k'$ in subband $k$, 
\begin{equation}
    \phi_{k,k'}[t] = e^{j2\pi k't}\sum_{m=-K}^{K}\tilde{\psi}[m]\psi[t+m]e^{-j\frac{2\pi}{K}m(k-k')},
    \label{eq:opt2}
\end{equation}
with $\tilde{\psi}[t]$ and $\psi[t]$ denoting the STFT analysis and synthesis windows with length $K$. Using the crossband filter $u_m(k,k',n)$ and the integer frame-delay $\delta_m^{\text{frame}}$, optimal TDOA compensation may be performed in the STFT-domain, equivalent to the optimal time-domain TDOA compensation filter $\underline{u}'_{m}[t]$ in \eqref{eq:time_domain}. \par
We now apply optimal TDOA compensation to the prediction-delayed microphone signal $x_m(k,n-\tau)$, where $\tau$ is the integer prediction delay of the reference microphone. This leads to a non-integer microphone-dependent prediction delay $\tau_m$ in the WPE algorithm (see Fig. \ref{fig:block_diag}), i.e.
\begin{equation}
\boxed{x_m(k,n-\tau_m) = \sum_{k' = 1}^{K}\sum_{l = -L_ {a}}^{L_{c}}x_m(k',n-l-\tau_m^{\text{int}})u_m(k,k',l),}
    \label{eq:crossband_taum}
\end{equation}
where $L_{a}$ and $L_{c}$ denote the number of acausal and causal taps in $u_m(k,k',n)$, and $\tau_m^{\text{int}} = \tau + \delta^{\text{frame}}_m$ denotes the integer component of the prediction delay $\tau_m$.
\subsection{Band-to-band approximation}
Since the crossband filter $u_m(k,k',n)$ in \eqref{eq:crossband_taum} is computationally complex and it has been shown in \cite{Avargel2007} that crossband filters do not necessarily lower reconstruction error in $x_m(k,n-\tau_m)$, we propose to use a band-to-band approximation of the optimal TDOA compensation in \eqref{eq:crossband_taum}, i.e.
\begin{equation}
    \boxed{x_m(k,n-\tau_m) = \sum_{l = -L_ {a}}^{L_{c}}x_m(k,n-l-\tau_m^{\text{int}})u_m(k,l),}
    \label{eq:bandtoband_taum}
\end{equation}
where $u_m(k,n) = u_m(k,k,n)$ denotes the band-to-band filter in \eqref{eq:opt1}.
\subsection{Integer prediction delays}
In order to further reduce computational complexity, we only consider the integer frame-delay component in \eqref{eq:bandtoband_taum} and ignore the band-to-band filter $u_m(k,n)$. This corresponds to applying coarse TDOA compensation to the prediction-delayed microphone signal $x_m(k,n-\tau)$, leading to an integer microphone-dependent prediction delay $\tau_m = \tau_m^{\text{int}}$, i.e.
\begin{equation}
    \boxed{x_m(k,n-\tau_m) = x_m(k,n-\tau_m^{\text{int}}).}
    \label{eq:integer_taum}
\end{equation}
\section{Experimental Evaluation}
In this section we evaluate the performance of the microphone-independent prediction delay and the proposed microphone-dependent prediction delays in the WPE algorithm for an acoustic sensor network with spatially distributed microphones, both using oracle as well as estimated TDOAs. In Section 5.1 we discuss the considered acoustic setup and the performance measures. In Section 5.2 we present the simulation results for the different TDOA compensation implementations proposed in Section 4.

\subsection{Acoustic setup and performance measures}
\begin{figure}[t!]
\centering
\includegraphics[width=0.55\columnwidth]{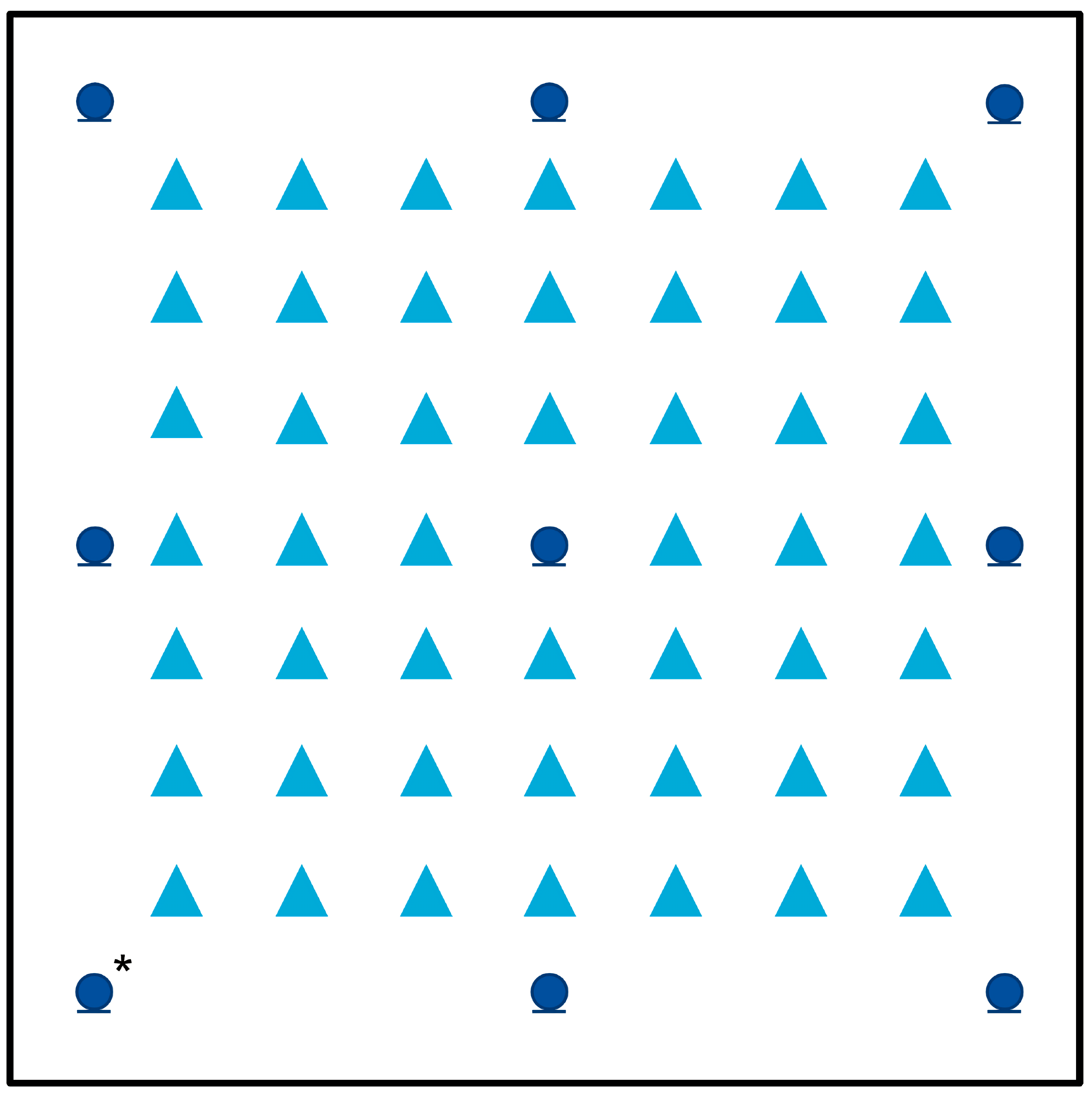}
\caption{Positions of $M=9$ spatially distributed microphones \includegraphics[width=0.2cm]{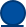} with fixed reference microphone (*) and 48 considered positions of single speech source \includegraphics[width=0.2cm]{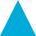}}
\label{fig:acoustic_scen}
\vspace{-0.4cm}
\end{figure}

We consider an acoustic sensor network with $M = 9$ spatially distributed microphones and a single speech source in a room of dimensions $8$m$\times8$m$\times5$m. Fig. \ref{fig:acoustic_scen} depicts the position of the microphones and the considered positions of the speech source inside the room. The microphones are placed on a $3\times3$ grid with equal spacing of dimensions $7.5$m$\times7.5$m at a height of $1.5$m. The reference microphone is chosen as the bottom-left microphone and fixed for all considered source positions. In total 48 positions of the speech source are considered on a $7\times7$ grid with equal spacing of dimensions $7$m$\times7$m at a height of $1.5$m, with one position of the speech source removed as it overlaps with a chosen microphone position. The minimum distance between any considered position of the speech source and the closest microphone is $0.5$m. \par
The reverberant microphone signals were generated by convolving a subset of anechoic speech from the TIMIT database \cite{TIMIT} with room impulse responses (RIRs) that were simulated using the randomized image method \cite{DeSena2015} with reverberation time $T_{60} \in \{500,750,1000\}$ ms. The signals were processed at a sampling rate of $16$ kHz using an STFT framework with frame size $K = 1024$ samples, frame shift $L_{\text{shift}} = 256$ samples and square-root Hann analysis and synthesis windows in \eqref{eq:opt2}. The WPE algorithm was implemented with prediction filter length $L_g \in \{8,12,16\}$ (proportional to $T_{60}$), prediction delay $\tau = 2$ and sparsity-promoting parameter $p = 0.5$ in \eqref{eq:sparsity_p}.
\par To estimate the TDOAs of the speech source between the microphones, we used the popular generalised cross-correlation with phase transform (GCC-PHAT) algorithm \cite{Knapp1976} with a frame size of 2048 samples (i.e. twice as large as WPE) and a frame shift of 1024 samples, since the frame size needs to be at least twice as large as the largest possible TDOA. \par
Dereverberation performance is evaluated using the frequency-weighted segmental signal-to-noise-ratio (FWSSNR), the perceptual evaluation of speech quality (PESQ) and the cepstral distance (CD) measure \cite{Kinoshita2016}. The reference signal used in these measures is the direct component in the reference microphone. The above measures are averaged across the 48 considered positions of the speech source.
\subsection{Simulation results}
\begin{figure}[H]%
\centering
\begin{subfigure}[H]{0.48\columnwidth}
  \includegraphics[width=\columnwidth]{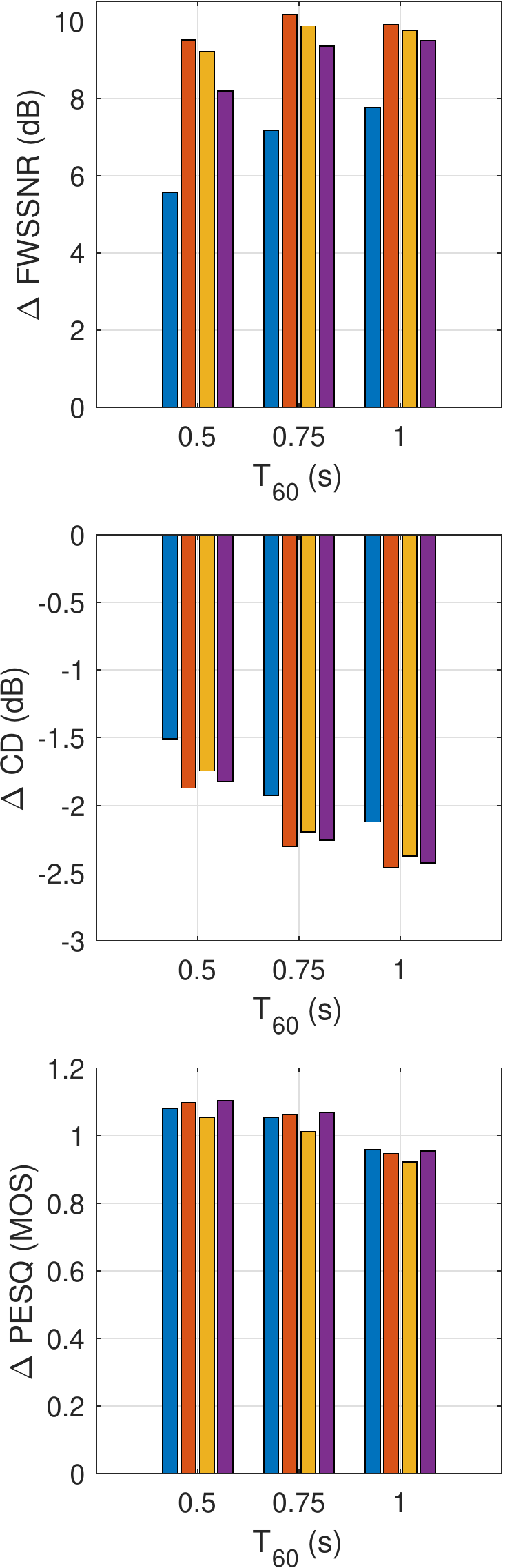}
  \vspace{-0.3cm}
  \caption{Oracle TDOAs}%
  \label{subfig: sdn}%
\end{subfigure}\hfill%
\begin{subfigure}[H]{0.48\columnwidth}
  \includegraphics[width=\columnwidth]{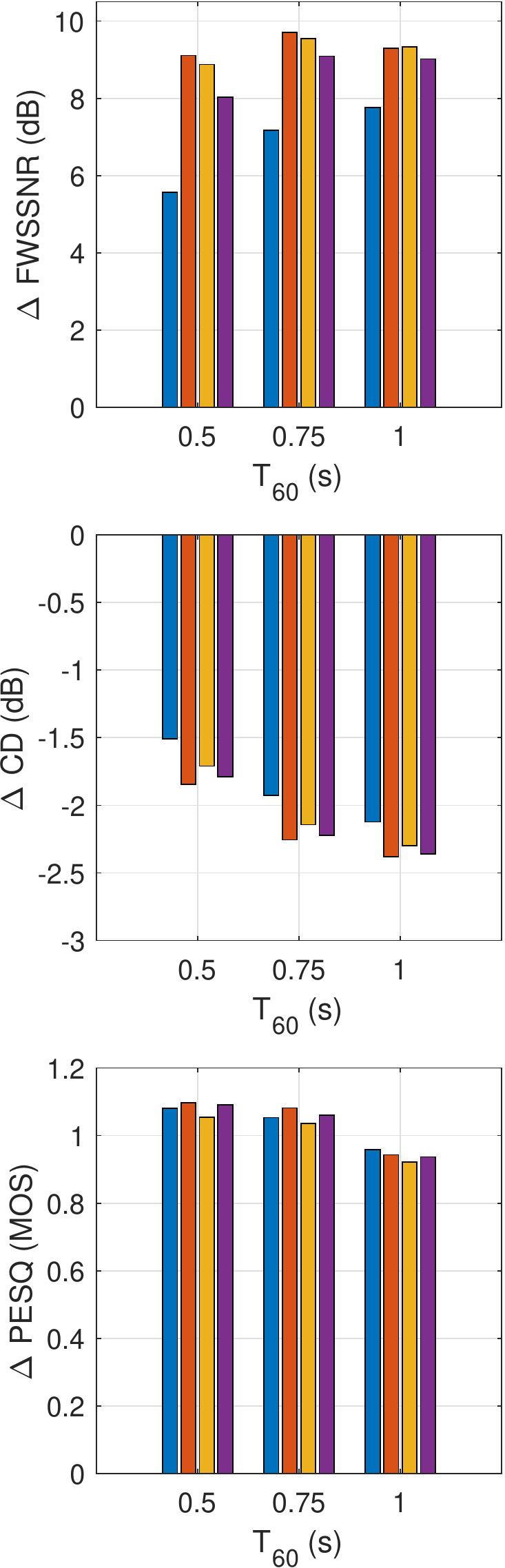}
  \vspace{-0.3cm}
  \caption{Estimated TDOAs}%
  \label{subfig: sdb}%
\end{subfigure}
\vspace{-0.2cm}
\caption{Average performance in terms of frequency-weighted segmental SNR improvement, CD improvement and PESQ improvement for all considered WPE algorithms for different reverberation times $T_{60}$, for (a) oracle TDOAs and (b) estimated TDOAs}
\label{fig:performance}
\vspace{-0.4cm}
\end{figure}

For the WPE algorithm, we consider the following prediction delays:
\begin{itemize}[leftmargin=0.4cm]
    \item MI: using a microphone-independent prediction delay $\tau$
    \item MD-NINT: using a microphone-dependent prediction delay $\tau_m$, performing optimal TDOA compensation using crossband filtering in \eqref{eq:crossband_taum}
    \item MD-NINT-B2B: using a microphone-dependent prediction delay $\tau_m$ with band-to-band approximation in \eqref{eq:bandtoband_taum}
    \item MD-INT: using a microphone-dependent integer prediction delay $\tau_m$ in \eqref{eq:integer_taum}
\end{itemize}
For all considered WPE algorithms, Fig. \ref{fig:performance} depicts the average performance for different reverberation times, both for oracle as well as for estimated TDOAs.
First, for oracle TDOAs it can be observed for all reverberation times that in terms of $\Delta$FWSSNR and $\Delta$CD the performance using microphone-dependent prediction delays is significantly better than using a microphone-independent prediction delay. In terms of $\Delta$PESQ, the performance is similar, although clear differences in speech quality are audible between using microphone-dependent prediction delays and using a microphone-independent prediction delay\footnote{Audio examples available on \\uol.de/f/6/dept/mediphysik/ag/sigproc/audio/dereverb/mdpd.html}. Comparing the different implementations of the microphone-dependent prediction delays, it can be observed that non-integer delays with crossband filtering (MD-NINT) clearly outperforms integer delays (MD-INT), while using the band-to-band approximation of non-integer delays (MD-NINT-B2B) is only marginally worse than using crossband filtering.
Second, using estimated TDOAs for TDOA compensation a very similar performance can be achieved as using oracle TDOAs for all considered WPE algorithms. This shows a high degree of robustness to estimation errors, even in highly reverberant environments.
\pagebreak
\begin{figure}[H]
\hspace{1.2cm}\includegraphics[width=0.8\columnwidth]{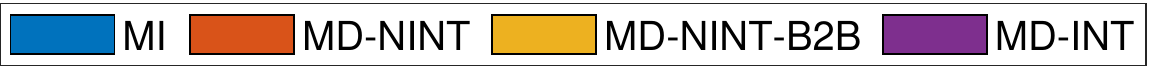}
\end{figure}%
\vspace{-0.6cm}
\section{Conclusion}
In this paper we have presented a WPE algorithm with microphone-dependent prediction delays. We have proposed to use these microphone-dependent prediction delays to perform TDOA compensation, which is especially relevant in acoustic sensor networks with spatially distributed microphones. We have considered several versions to perform TDOA compensation in the STFT-domain, i.e. optimal TDOA compensation with non-integer delays using crossband filtering, approximate TDOA compensation using band-to-band filters or coarse TDOA compensation with integer-frame delays. The experimental evaluation showed that using microphone-dependent prediction delays to perform TDOA compensation improves the performance compared to a microphone-independent prediction delay, where the best dereverberation performance is achieved using non-integer delays with crossband filtering closely followed by the computationally less complex band-to-band approximation. Investigating the performance of the proposed schemes for acoustic scenarios with additive noise, a moving source or multiple sources are directions for future research.




\pagebreak
\bibliographystyle{IEEEbib}
\balance
\bibliography{refs}
\end{document}